\documentclass[conference]{IEEEtran}

\usepackage{epsfig}
\usepackage{graphicx}
\usepackage{amsmath}
\usepackage{amssymb}
\usepackage{amsxtra}
\usepackage{here}
\usepackage{rawfonts}
\usepackage{times}
\usepackage{url}
\usepackage{cite}
\usepackage{rotating}
\usepackage{multirow}

\usepackage{adjustbox}

\usepackage{slashbox}
\usepackage{epstopdf}

\usepackage{cite}

\newlength{\aligntop}
\newlength{\alignbot}
\makeatletter

\IEEEoverridecommandlockouts

\begin{document}
\title{\huge Smart Grid Security: Threats, Challenges, and Solutions\vspace{-0.55cm}}
\author{\IEEEauthorblockN{Anibal Sanjab$^1$, Walid Saad$^1$, Ismail Guvenc$^2$, Arif Sarwat$^2$, and Saroj Biswas$^3$} \IEEEauthorblockA{\small
$^1$ Wireless@VT, Bradley Department of Electrical and Computer Engineering, Virginia Tech, Blacksburg, VA USA,\\
 Emails: \url{{anibals,walids}@vt.edu}\\
 $^2$ Department of Electrical and Computer Engineering, Florida International University, Miami, FL, USA, Email: \url{{iguvenc,asarwat}@fiu.edu}\\
 $^3$ Department of Electrical and Computer Engineering, Temple University, Philadelphia, PA, USA, Email: \url{saroj.biswas@temple.edu}\vspace{-0.4cm}\\
 }%
%\thanks{This research was supported by the U.S. National Science Foundation under Grant CNS-1446621.}
    }
\date{}
\maketitle

\begin{abstract}
The cyber-physical nature of the smart grid has rendered it vulnerable to a multitude of attacks that can occur at its communication, networking, and physical entry points. Such cyber-physical attacks can have detrimental effects on the operation of the grid as exemplified by the recent attack which caused a blackout of the Ukranian power grid. Thus, to properly secure the smart grid, it is of utmost importance to: a) understand its underlying vulnerabilities and associated threats, b) quantify their effects, and c) devise appropriate security solutions. In this paper, the key threats targeting the smart grid are first exposed while assessing their effects on the operation and stability of the grid. Then, the challenges involved in understanding these attacks and devising defense strategies against them are identified. Potential solution approaches that can help mitigate these threats are then discussed. Last, a number of mathematical tools that can help in analyzing and implementing security solutions are introduced. As such, this paper will provide the first comprehensive overview on smart grid security.            
\end{abstract}

\section{Introduction}\label{sec:Intro}
\begin{figure}[t!]
  \begin{center}
   %\vspace{-0.35cm}
    \includegraphics[width=9cm]{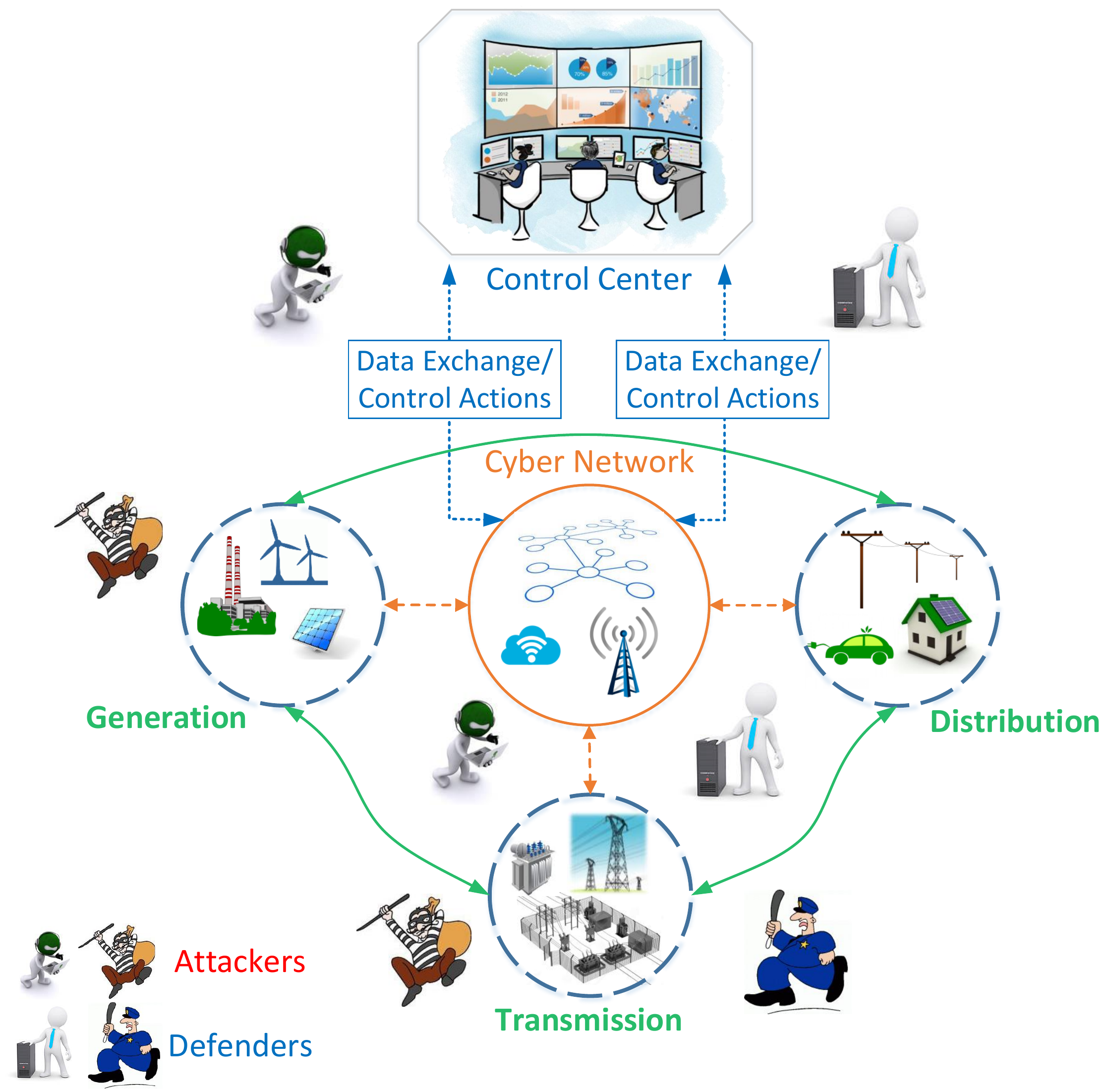}
    %\vspace{-0.55cm}
    \caption{\label{fig:TheSmartGrid} Illustration of a smart grid architecture highlighting the underlying cyber layer and security threats.}
  \end{center}%\vspace{-0.9cm}
\end{figure}

Realizing the vision of a smart electric grid is contingent upon the effective integration of new information and communication technologies into the traditional generation-transmission-distribution physical systems. This, in turn, will give rise to a cyber-physical power system known as the \emph{smart grid (SG)} in which a cyber layer that handles computations, communication, and data exchange, is tightly coupled with the physical system which handles the generation, transmission, and distribution of electric power as shown in Fig.~\ref{fig:TheSmartGrid}.
 
Despite the significant advantages introduced by this cyber-physical coupling, the resulting dense interconnectivity between various SG elements and the increased reliance on its cyber system makes the grid more vulnerable to a multitude of cyber-physical attacks (CPAs) that aim at compromising its functionalities. This increased SG vulnerability is corroborated by the recent discovery of the control system malware, known as Stuxnet~\cite{Stuxnet}, which targets programmable logic controllers (PLCs) of industrial systems giving the adversary the ability to have control over the physical system.  
The fear of such threats has peaked after the first CPA-induced Ukranian blackout which has recently affected 225,000 customers spanning several Ukranian cities~\cite{NERCUkranian}. 
Clearly the security of the smart grid as one of the most critical technical challenges facing its deployment~\cite{MITReportPerGridSec}.

With such culminating risks, %vulnerability analysis as well as preventive, mitigative, and restorative measures and policies must be devised to secure the grid and maintain its availability.
%To this end, 
identifying and understanding potential threats which can target the SG is essential to achieve a more secure grid. This identification enables devising new security measures and strategies to thwart such attacks and make the grid more robust and resilient. However, due to the complex and large-scale nature of the SG, various challenges accompany these diagnostic and corrective efforts. 

The main contribution of this article is to provide the first comprehensive overview on the security threats facing the SG. In particular, the goal of this paper is threefold: 
\begin{enumerate}
\item{identify and explore the CPA threats which can target the smart grid,} 
\item{discuss the unique challenges of analyzing SG security problems, and,}
\item{propose solution approaches and analytical frameworks which can help in analyzing the security of the SG and devising proper defense strategies.}
\end{enumerate} 

In a nutshell, due to its necessity for all facets of daily life activities, interrupting the supply for electricity emerged as a lucrative target for adversarial activities. Insuring the sustainability and availability of electricity supply via SG functions is imperative but challenging as it requires a comprehensive knowledge of the various aspects of SG security. To this end, this paper aims at providing an inclusive investigation of the various security threats, challenges, and solutions pertaining to SG security. %Due to the necessity of With the recent culminating threats targeting the SG, understanding all aspects of SG security is essential to devise informed and adequate solution. 
                                                                                                                                
\section{Smart Grid Security Threats}\label{sec:SecThreats}
Given its cyber-physical nature, %the security vulnerabilities of the SG stem from the cyber layer, its accompanying physical system, and the tight cyber-physical interconnections between them. Naturally, 
the SG will inherit physical and dynamic system threats as well as well-known communication and network threats such as those targeting its integrity and availability\footnote{Privacy is yet an important facet of smart grid security that is out of scope of this article due to space limitations.}. %, and privacy. 
However, the goals of such attacks will significantly differ from those sought by adversaries targeting classical cyber systems, such as communication networks. 
For instance, an attack on the SG primarily aims to disturb the operation of the physical generation, transmission, and distribution systems by exploiting their reliance on their underlying cyber layer. This is in contrast to conventional network attacks which typically seek to solely cause some sort of damage to or interruption of the cyber layer's functionality. %For example, an attack using which an eavesdropper is able to extract data sent by a set of measurements units is a cyber security breach. However, this breach does not constitute a significant threat to the operation of the cyber-physical power grid unless the attacker uses such information to improve its knowledge of the system in order to design a future wide scale attack. On the other hand, eavesdropping in a cyber network used for wireless communication is significant since it entails unauthorized access to user's private information. 
%
%In addition to integrity and availability threats, other specific types of attacks can also target the grid's physical and control systems. 
%Having clarified the unique features of smart grid security, next, we expose various cyber-physical threats which can target the grid while focusing on understanding and quantifying their effects on the stability and operation of the system. %rather than diving into the technicalities of how these attacks can be technically carried out.    
The key SG security threats are discussed next.
\subsection{Integrity}\label{subsec:Integrity}
Integrity refers to the credibility of the data collected and transferred over the grid. Attacks that target this integrity can cause false estimation of the real-time state of operation of the system as well as lead to the unobservability or even instability of the system. Next, we present two key types of integrity attacks.

\subsubsection{Data injection attacks (DIAs)}\label{subsubsec:DataInjection} 
DIAs consist of an adversary manipulating exchanged data such as sensor readings, feedback control signals, and electricity price signals. Such attacks can be done by compromising the hardware components (as in the case of Stuxnet), or intercepting the communication links. The most studied type of DIAs is the one that targets the grid's state estimator. The states of a power system consist of the voltage magnitudes and phase angles at every bus. Estimation of these states %and the topology of the system %(along with the corresponding transmission line series/shunt impedances/reactances) 
enables complete monitoring of the power and current flows throughout the grid. %of the currents as well as real and reactive powers flowing throughout the grid. 

%In practice, since these states cannot be measured at every location and to provide redundancy to phase out outliers, 
To estimate these states, a number of measurements are collected from around the grid %and used to estimate the states. Such measurements
which include %quantities such as
real and reactive power flows over transmission lines or injected/withdrawn at buses, voltage magnitudes, and, %throughout the system whereas the deployed PMUs, which have played an essential role in the rise of SGs, have the capabilities to measure, in a 
due to the placement of phasor measurement units (PMUs), %synchronized manner, 
synchronized voltage and current phase angles. % at the various substations. 
%Since voltages, currents, and powers are interdependent quantities, 
%The collection of these measurements can be used to estimate the voltage magnitudes and angles across the system. 
%
%In this respect, 
The collected measurements are then fed to a state estimator which, using a maximum likelihood estimator, generates real-time estimates of the states which are used for operational and pricing purposes. 

Therefore, manipulating the collected measurements results in a false estimate of the state of operation of the system. In turn, such false states can lead to incorrect operational actions whose effects can range from inducing incorrect pricing to destabilizing the power system. In practice, a bad data detection (BDD) mechanism is deployed to detect outliers. However, the authors in~\cite{liu1stdatainjection} showed the existence of a \emph{stealthy data injection attack model} that can manipulate the state estimation outcome without being detected by common BDD mechanisms. The effects of such attacks vary widely depending on the goal of the attacker. 
\begin{itemize}
\item{\emph{System damage:} Some DIAs are primarily concerned with damaging the system and have a purely destructive nature, such as in terrorist attacks. 
For example, an attacker can manipulate system measurements so that a congested transmission line falsely seems to not have reached its thermal transmission limit. Thus, based on these false estimates, %, not knowing that this line has already reached its thermal limit, 
the system operator %can decide to route power along this transmission line, to reduce stress on other parts of the system, 
would route more power over the line which leads it to over-heat and sag. This sagging reduces the distance in between the line and ground (or other objects in between) %due to thermal expansion , 
%thus reducing its distance with objects over the line path, 
which can trigger a line to ground fault. Under stress conditions, such a fault can induce large fluctuations in system dynamics that can lead to tripping additional lines, disconnecting generators, load shedding, or even a system blackout.}       
\item{\emph{Financial benefit:} %rather then aiming at damaging the system, an attacker might be a market participant who uses DIAs to manipulate electricity prices and reap financial benefit. 
Real-time pricing in electricity markets is based on %energy offers and demand bids submitted by generation owners (GENCOs), load serving entities (LSEs), and energy brokers, and on the 
the state estimator's real-time estimate of the state of operation of the system. Thus, corrupting the state estimates using DIAs leads to manipulation of the electricity prices in a way that is lucrative to the attacker~\cite{SanjabSaadDataInjTSG}. 
\begin{figure}[t!]
  \begin{center}
   %\vspace{-0.35cm}
    \includegraphics[width=9cm]{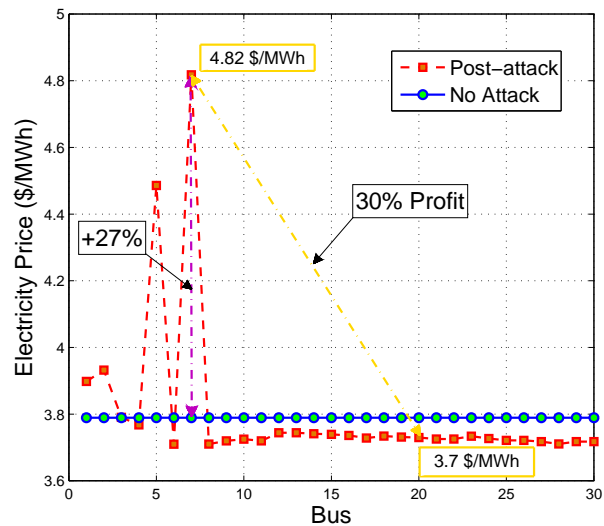}
    %\vspace{-0.55cm}
    \caption{\label{fig:LMPswDifferentAttacks} Electricity price manipulation using data injection attacks.}
  \end{center}%\vspace{-0.9cm}
\end{figure}

This is illustrated in Fig.~\ref{fig:LMPswDifferentAttacks} which corresponds to a DIA targeting measurements of three line flow measurements %units over lines 5, 9, and 11 
over the IEEE 30-bus test system. 
Clearly, without any attack, the prices are equal throughout the system. However, the attack successfully manipulated the prices causing, for example, a $27\%$ increase in the price at bus $7$. Thus, suppose that the attacker is a market participant who sells power at bus $7$ and buys the same amount of power at bus $20$. As seen in Fig.~\ref{fig:LMPswDifferentAttacks}, without any attack, this transaction generates no profit. However, in the presence of an attack, the participant reaps a $30\%$ profit from this transaction.} 
\end{itemize}

Beyond targeting the state estimator, data injection attacks can target wide area protection, monitoring, and control (WAPMC) schemes %. %the output of sensors as well as the control signals used for stabilizing the dynamic power system. In fact, power system stabilizers rely on local and global measurements. Thus, manipulating these measurements or control signals can lead to destabilizing the system. 
%Another area that can be vulnerable to data injection attacks is wide area protection, monitoring, and control (WAPMC). WAPMC schemes 
which rely on global data collected from around the system to detect the occurrence of a disturbance and take corrective actions to stop its propagation. In this regard, manipulating the exchanged data can lead to a false characterization of a disturbance leading to false disconnection of lines, generators, or loads~\cite{SanjabSaadCPSWeek}.  

\subsubsection{Time synchronization attacks}\label{subsubsec:TimeSync}
%PMUs have revolutionized real-time sensing and monitoring of electric transmission systems and have played a crucial role in wide area monitoring, protection and control~\cite{WideAreaPhadke}. PMUs 
To better monitor the grid, there has been an increased use of PMUs -- high-speed measurement units (typically 30-60 samples/second) capable of measuring the voltage and current phasors as well as local frequencies. % at the locations at which the they are placed. %PMUs collect 30-60 samples per second which is much higher than a standard 0.5-1 samples per second supervisory control and data acquisition (SCADA) system. Moreover, 
Given that the measurement devices are spread around the system, sending their collected measurements to data concentrators or control centers is subject to transmission delays. Therefore, in order to properly align and analyze the measurements, all the collected PMU data are synchronized based on a time reference provided by a global positioning system (GPS) signal. This time referencing provides a time stamp to each collected measurement. The high speed sampling capabilities and, most importantly, the synchronization between the collected measurements enable accurate real-time wide area monitoring, protection, and control of the SG. %as well as global protection against faults~\cite{WideAreaPhadke}. 

Here, an adversary can manipulate the time reference of the time stamped measured phasors to create a false visualization of the actual system conditions thus yielding inaccurate control and protection actions. Attacks that target PMU time synchronization are known as \emph{time synchronisation attacks (TSAs)}~\cite{TimeSynchroAtt}. Using TSAs, the GPS signal is spoofed and counterfeited by the attacker so that PMU sampling is done at the wrong time hence generating measurements with wrong time stamps. Recent results in~\cite{TimeSynchroAtt} %has studied the effects of TSAs on transmission line fault location detection and voltage stability monitoring. 
%
%The results 
have shown that TSAs can produce significant fault location errors which can go up to $180$ km for a line of length $400$ km and even trigger a false alarm regarding the presence of a fault. This false alarm can result in a disconnection of a transmission line which can then trigger a cascading chain of failures across the grid. 
Such a false disconnection was one of the main culprits that led to the  North American Northeast blackout in 1965~\cite{SanjabSaadCPSWeek}.%~\cite{TheGridSchewe2007}.
%For example, the North-American Northeast blackout of 1965, which left around 30 million people in Ontario, Canada and 8 U.S states with no power for up to 13 hours, was caused by a false disconnection of a transmission line due to incorrect setting of a protective relay~\cite{TheGridSchewe2007}. %human error (incorrect setting of a protective relay by maintenance personnel)~\cite{TheGridSchewe2007}.
%
%When it comes to voltage stability monitoring, PMUs are used to collect voltage and current phasor measurements which are used to calculate what is known as voltage stability indicators. The authors in~\cite{TimeSynchroAtt} have shown that TSAs can lead to disabling the voltage instability alarm, in the case of occurrence of disturbance, or to falsely implement wrong voltage stabilization measures.   
%
%\subsubsection{Other Dynamic System Attacks}\label{subsec:DynamicThreats}
%An additional type of attacks that can target the smart grid is known as \emph{replay attacks} (RAs). In RAs, the adversary, aims at injecting input data in the system without causing changes to the system's measurable outputs. To launch this attack, an adversary compromises sensors, monitors their outputs, learns from them, and repeats them while injecting its attack signal. The effects of RAs on the smart grid have been investigated in~\cite{ReplayAttacksSG}. 
 
\subsection{Availability}\label{subsec:Availability}
Availability pertains to the accessibility to every grid component as well as to the information transmitted and collected, whenever needed. Attacks compromising this availability are known as \emph{denial of service (DoS)} attacks that can block key signals to compromise the stability of the grid and observability of its states. 

\begin{figure}[t!]
  \begin{center}
   %\vspace{-0.35cm}
    \includegraphics[width=9cm]{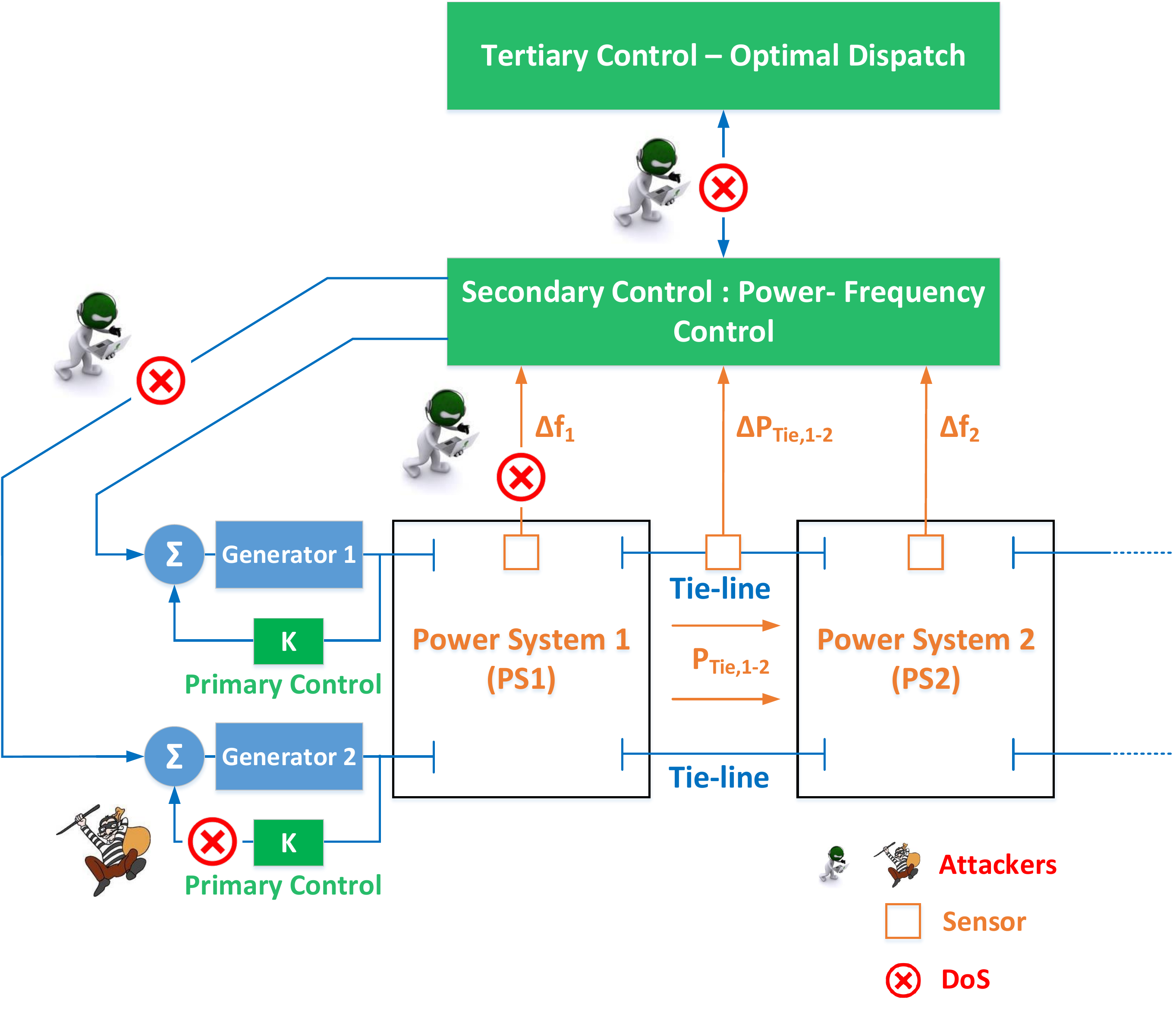}
    %\vspace{-0.55cm}
    \caption{\label{fig:ThreeLayerControlSG} Illustration of the three-layered control architecture of a smart grid: $\Delta f_i$ is the deviation in frequency within PS$_i$ from the nominal frequency (60 Hz in most of North and South America, 50 Hz in most other areas of the world), and $\Delta P_{\textrm{tie,1-2}}$ is the deviation in power flowing over the tie-lines from the scheduled value. Generators 1 and 2 are part of PS$_1$.}
  \end{center}%\vspace{-0.9cm}
\end{figure}

In this regard, maintaining generation-load balance is essential for the SG operation. Indeed, %based on Newton's second law, 
the angular frequency of a synchronous generator %the acceleration of a synchronous generator's rotor 
is based on the difference between the electric power it serves and the input mechanical power of its turbine provided by, for example, burning fuel or coal. For the generator to retain constant angular frequency, its mechanical power should always match its connected electric load. Thus, if the mechanical input is kept constant, an increase in the electric load leads to %the deceleration of the generator's rotor 
a drop in angular frequency while a decrease in load leads to a rise in frequency. 
Consequently, generators are equipped with local and global control systems that typically follow a three-layer design as shown in Fig.~\ref{fig:ThreeLayerControlSG} to maintain a constant frequency. 

The \emph{first layer} consists of a local primary proportional controller which aims at adjusting the mechanical power of the generation to match the changes in electric load. This proportional control reduces the frequency deviation but maintains a steady state error. To eliminate this steady state error, a central integral control is used in which various generators participate. This integral control represents the \emph{secondary control layer}. 
The \emph{tertiary control}, corresponds to a supervisory control responsible for allocating enough spinning reserve %, needed by the primary control, 
and for optimally dispatching the units participating in the secondary control. 
Moreover, the SG consists of an interconnection of many power systems which are connected using tie-lines. Thus, a deviation in frequency at one system triggers deviations in the power flowing over the tie-lines. Consequently, a control scheme known as \emph{power-frequency control} regulates the tie-line power flow based on the sensed frequency deviations. %is a generation control which regulates the power flowing in between various systems to reduce deviations from the scheduled power exchange level. This is achieved using the combination of primary and secondary controls. Performing this power-frequency control while minimizing the associated generation cost consists of what is known as automatic generation control (AGC) and involves all three layers of control. 

Therefore, the dynamic stability of the grid is crucially dependent on the availability of the sensor measurements and control signals provided by the three layers of control. In fact, using DoS to block the primary control signal and prevent it from decreasing the mechanical input power following a drop in the load leads to the acceleration of the generator and its shut down by an over-frequency relay. Moreover, a DoS blocking secondary control from eliminating steady state frequency errors can lead to the loss of synchronism between generators. Analogously, blocking tertiary control can have similar effects and can lead to suboptimal operation incurring large monetary losses.  

This hence also sheds light on the difference between threats on the SG as a CPS and threats pertaining to conventional cyber systems. In fact, compromising availably can destabilize the SG. In contrast, a DoS may not, in most cases, stop the operation of a cyber system, but it will typically incur delays.  
 %In fact, as can be seen in Fig.~\ref{}, the nominal frequency cannot be tota           

%Compromising availably can hence destabilize a cyber-physical system in the contrary to the case of conventional networked systems in which a temporary DoS may not, in most cases, lead to a stop in the operation of the system~\cite{SecCtrlBerkeley}.

\subsection{Additional Dynamic System Attacks}\label{subsec:DynamicThreats}
As a dynamic control system, various dynamic system attacks (DSAs) can be launched at the SG. One well investigated such type of attacks is known as \emph{replay attacks (RAs)} which can have serious effects on system stability~\cite{ReplayAttacksSG}. In RAs, the adversary injects input data in the system without causing changes to the measurable outputs. To launch this attack, an adversary compromises sensors, monitors their outputs, learns from them, and repeats them while injecting its attack signal. %The effects of RAs on control system performance is investigated in~\cite{ReplayAttacksSG}. 
Another type of DSAs is known as \emph{dynamic data injection attacks (D-DIA)} which uses knowledge of the grid's dynamic model to inject data that causes unobservability of unstable poles~\cite{AttDetDorfler}. As a result, a successful D-DIA prevents the grid's operator from detecting instability which, in turn, can lead to a system collapse. 
%knowing that the system has become unstable and hence no corrective actions would be taken to prevent the system from collapsing. 
A \emph{covert attack} is one other type of DSAs that is basically a closed loop version of an RA~\cite{AttDetDorfler}.    

\subsection{Physical Threats}\label{subsec:Physical}
Given the wide footprint over which the power system is physically spread, the danger of physical attacks in which an adversary physically attacks a physical component such as a generator, substation, or transmission line is prominent. 
For example, components can be physically attacked remotely using a rifle as was the case in a sniper attack which targeted a substation in California in 2013~\cite{SniperSGAttack}. % which almost caused a blackout in parts of the Silicon Valley. 
Another type of physical attacks consists of physical manipulation of smart meters for energy theft purposes. 
%
%\subsection{Privacy}\label{subsec:Privacy}
%Due to the human participation in the smart grid at the distribution level as interactive customers and at the transmission level as energy bidders, privacy will always remain a concern. In fact, various studies have focused on analyzing privacy concerns when it comes to readings of smart meters and energy consumption patterns. In fact, from smart meters' readings, private information such as which appliances are being used at which time, whether an individual is present in its dwelling or not, and the type of activities that a person is performing can be extracted~\cite{PrivacySmartMeter}.

\subsection{Coordinated Attacks}\label{subsec:CoordinatedAtt}
The power system typically incorporates robustness measures that helps it survive potential failures. Under typical system conditions, an attack leading to the failure of one or few components might not always have significant effects on the grid's operation. For example, the power system follows the so-called ``$N-1$'' security criterion which instills redundancies in the system design allowing the preservation of the system's state of normal operation even after the loss of one of its $N$ components. 

However, \emph{coordinated attacks (CAs)} can still be launched by resourceful adversaries that exploit the dense interconnections between grid components %, provided by the cyber layer, 
to launch simultaneous attacks of different types 
%(data injection, DoS, TSAs, RAs), concurrently 
targeting various components. For example, the recent CPA-caused blackout of the Ukranian grid is a CA which concurrently targeted three power distribution companies. The adversary compromised a number of their computers to gain control of the SCADA system to simultaneously disconnect around $27$ substations~\cite{NERCUkranian}. %while, at the same time, launching a DoS against the power companies' call centers~\cite{NERCUkranian}.      

CAs are the most challenging types of attacks since they can surpass traditional reliability and robustness design solutions and require a multi-layered security solution approach. %The challenges pertaining to devision security solutions for the SG are introduced next.

Table~\ref{Tab:CPSSecThreats} summarizes the discussed threats, their associated security types, and their main potential SG targets.  
\begin{figure*}[t]
\centering
\begin{table}[H]
	\caption{\large{SG Security Threats}}\label{Tab:CPSSecThreats}
	\begin{center}
		\begin{tabular}{|c|c|c|}
		\hline
			   \textbf{Threat Label} & \textbf{Security Breach Type} & \textbf{Main SG Target}  \\ \hline
                   Data injection attacks (DIA) & Integrity & State estimator, WAPMC \\ \hline
                   Time synchronization attacks & Integrity & PMUs, WAPMC \\ \hline
                   Denial of service attacks & Availability & Primary, secondary, and tertiary controls, WAPMC \\ \hline
                   Dynamic system attacks & Dynamic integrity & Primary, secondary, and tertiary controls\\  
                   (Dynamic DIA, Replay, Covert) & &\\ \hline                 
                   %Replay Attack & Integrity & Primary, secondary, and tertiary controls\\ \hline
                   %Covert Attack & Integrity & Primary, secondary, and tertiary controls\\ \hline
                   Physical destruction & Physical & Physical system components \\ \hline
                   Meter manipulation & Physical & Smart meters \\ \hline
                   Coordinated attacks & All of the above & All of the above \\ \hline                  
		\end{tabular}
	\end{center}
\end{table}  
%\\\bottomrule %\nonumber
\end{figure*}
%\begin{turn}{90} $\>\>\>$Integrity$\>\>\>$ \end{turn} 
\section{Challenges}\label{sec:Challenges}
The aforementioned threats naturally give rise to a number of key challenges, as detailed next.
\subsection{Limitation of Traditional Cyber Security Solutions}\label{subsec:CyberSec}
Existing cyber security solutions that have been devised for cyber systems are invaluable for improving SG security. For example, existing intrusion detection and cryptographic solutions will certainly contribute towards better grid security. However, some of these solutions might not be directly applicable to the SG due to a number of reasons:
\begin{enumerate}
\item{\emph{Presence of a physical system:} existing cyber security solutions do not consider the presence of a physical system. However, as discussed in Section~\ref{sec:SecThreats}, one of the main goals of attacks on the SG is to damage to the physical system. Thus, SG attacks are designed based on the physical effects that they can cause. Therefore, any security solution that does not directly account for the physical system is simply not adequate for defending the smart grid.}
%\item{Inadequate solution complexity: cryptographic algorithms typically requires a high energy overhead which might not always be available at some sensors or metering devices as for example protection relays, smart meters, etc.} 
\item{\emph{Risk management and diffusion:} the analysis of the propagation of attacks in a physical system is different from that corresponding to cyber systems. For instance, the study of how computer malware propagates in a cyber system is different from how failures cascade and propagate throughout a dynamic CPS such as the SG.}
%\item{\emph{Human-in-the-loop:} humans play a very essential role in the smart grid at the distribution as well as generation and transmission levels. As a result, cyber security solutions designed for a certain type of human interactions might not be perfectly applicable to the case of smart grids.}
\end{enumerate}
\subsection{Limitations of Existing Reliability Evaluation Solutions}\label{subsec:Reliability}
There exists several studies for improving the power system's reliability and availability via various means such as improving redundancies and maintenance processes. 
%availability by reducing the probability of failure of its components, improving maintenance processes and shortening repair times, including redundancies to decrease the global failure probabilities, and testing for the existence of hidden failures. 
Such existing reliability designs are primarily concerned with studying failure events which are likely to occur up to a certain level. However, a CA can cause various coordinated failures, as discussed in Section~\ref{subsec:CoordinatedAtt}, that have very low probability of naturally happening and, as such, they are not typically accounted for in reliability analyses. %For instance, as a reliability improvement measure making the grid more robust against failures, a power system, through what is known as the "$n-1$ security criterion" (allowing the system to preserve its state of normal operation even after the loss of one of its $n$ components) can possibly withstand the occurrence of a failure at a particular location; however, it cannot withstand the occurrence of coordinated and simultaneous failures. In fact, in response to the blackout that Brazil experienced in November of 2009, which is one of the largest blackouts in history depriving 70 million people of electric power and affecting more than half the states in Brazil and all of Paraguay, the director general of the Brazilian National Electricity System Operator (ONS) stated that "no system is designed to support the planning ahead of contingencies with such extremely remote probabilities".  
%However, undeniably, in standard reliability evaluation analysis, coordinated failures are very hard to naturally happen and hence cannot be completely accounted for. While on the other hand, an adversary can, based on the heavy interconnection between the control system and its underlying communication and computation backbone, coordinate an attack putting a control system into worst case scenarios, that it was not designed to withstand, leading to its failure. Since designing a system that is 100\% reliable is technically impossible, and since the increasing number of interconnected, non-redundant, components leads to a reduced reliability levels, securing such tightly interconnected cyber-physical systems is highly challenging. 
%
In fact, the cyber layer has provided an increased reachability for the adversaries using which various components that are % the various components of the power system. To this end, in a traditional power system setting, it might have not been possible for an adversary to launch simultaneous physical attacks on different sites 
located at separated geographic locations can be concurrently targeted. %A threat that is unlikely in a traditional power systemHowever, with the interconnection provided by the cyber layer, such simultaneous attacks could be possibly launched. 
\subsection{Limitations of Conventional Control-Theoretic Solutions}\label{subsec:Reliability}
Conventional control-theoretic security analyses are primarily concerned with designing robust controls which can preserve operational requirements in face of exogenous disturbances. However, such analyses do not explicitly account for the cyber layer and all the underlying cyber threats that it can introduce to a CPS such as the SG.
%\subsection{High Complexity}\label{subsec:complexity}
%The power system is known to be one of the most complex dynamic systems ever built. Moreover, the integration of communication and information technologies into the power system has increased its interconnectivity and interdependencies which has led to even more complexity. Thus, deriving security measures and policies is continuously faced with the challenge of fully understanding and accurately modeling the interconnectivity and dynamics of the complex cyber-physical smart grid. 
\subsection{Tradeoff Between Security and Performance}\label{subsec:SecvsPerformance}
SG security solutions must be inherently cognizant of the performance of the system. In particular, these solutions must seamlessly integrate with the grid with minimal disruption to its operation and performance. This tradeoff between security and performance is much more pronounced in the SG, compared to communication networks, due to the CPS nature of the grid. For example, the best security strategy to thwart cyber attacks from penetrating the grid is to completely eliminate wireless and Internet connectivity. Even though this will improve the security of the grid, it will deprive it from all the economic and operational advantages that the cyber layer introduces. Thus, finding the best security strategy while meeting stringent performance requirements is a key challenge. 
  
\subsection{Aging Components}\label{subsec:AgingSystem} 
One of the reasons that renders SG security even more challenging is that most of the components of the system were designed and implemented decades ago. Hence, at the time of their design, cyber layer integration had not been proposed yet; and thus, the security threats that it introduces had not been anticipated. Therefore, patching mechanisms to accommodate for the newly introduced threats are of high importance and, undeniably, pose serious challenges.
%
%\subsection{Large Investment Requirements}\label{subsec:Investment}
%Patching and securing the smart grid require very large investments due to the huge number of components that it includes. Allocating such investments to upgrading the security of the smart grid is usually very challenging. Hence, security solutions implementation is usually bounded by tight budgetary constraints which make meeting the intended security requirements even more complex.  
%  
\section{Solution Approaches}\label{sec:Solution}
\begin{figure}[H]
  \begin{center}
   %\vspace{-0.35cm}
    \includegraphics[width=9cm]{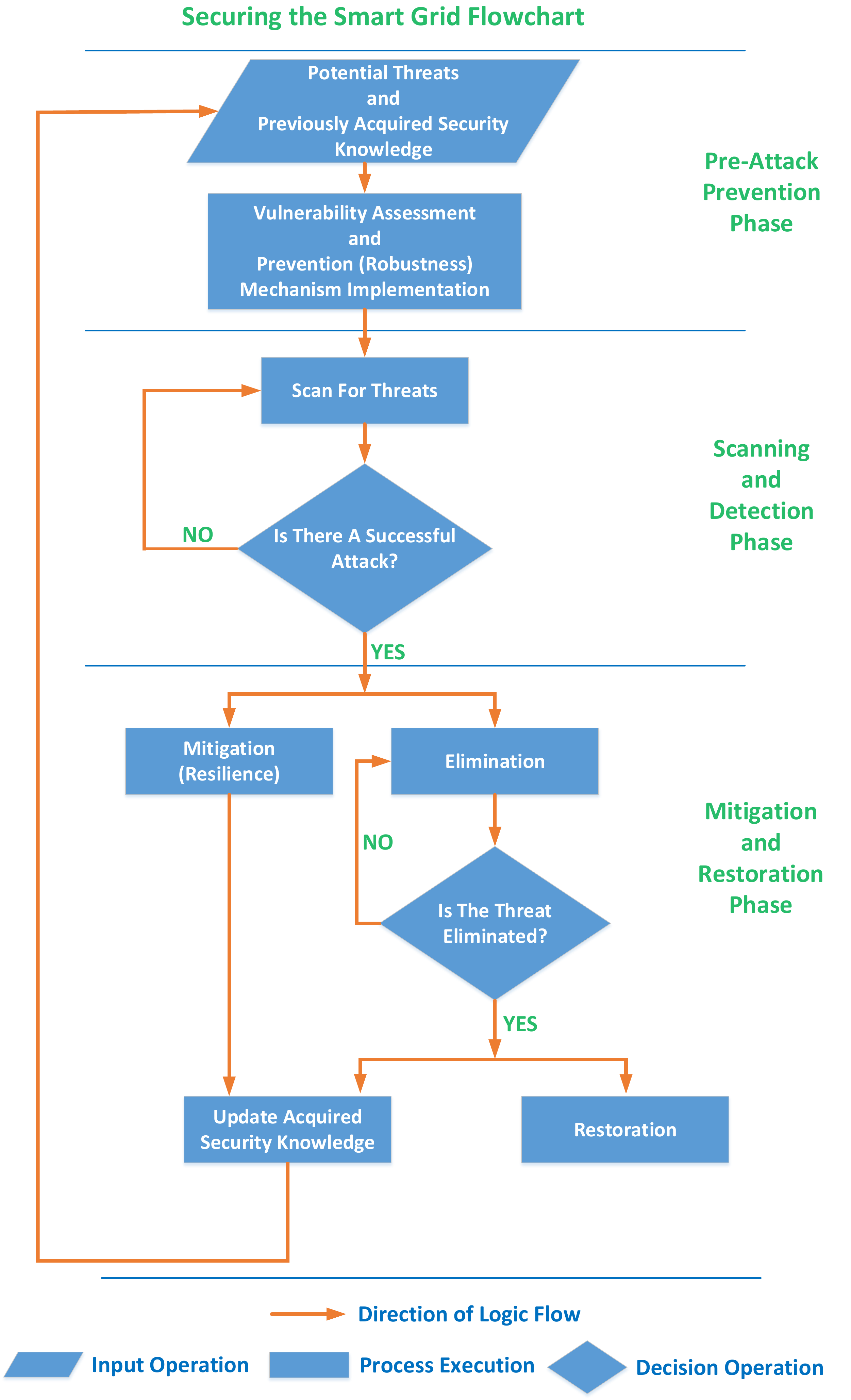}
    %\vspace{-0.55cm}
    \caption{\label{fig:SolutionApproach} Flowchart illustrating a systematic approach for the defense of the grid against CPAs.}
  \end{center}%\vspace{-0.9cm}
\end{figure}

Given the culminating threats facing the SG, security solutions must be developed to thwart potential attacks and maintain the operation of the grid. 
In particular, due to the high complexity involved, a systematic approach for securing the system is needed. To this end, a proposed systematic approach is illustrated in Fig.~\ref{fig:SolutionApproach} and is detailed next.  
\subsection{Prevention Phase}\label{subsec:Prevention}
The prevention phase involves reinforcing the security of the system to prevent any attack from successfully intruding and intervening in its operation. Here, various types of analyses can be performed:
\begin{enumerate}
\item{\emph{Vulnerability assessment and risk management:} vulnerability assessment consists of determining which grid components are vulnerable to which types of threats. Using past data, the SG operator can identify which components have historically been subject to which attacks. For example, since spoofing attacks targeting GPS signals are common, PMUs are expected to be vulnerable to spoofing and TSAs~\cite{TimeSynchroAtt}. Moreover, the vulnerability of some components can be also analyzed analytically and experimentally. For example, a non-encrypted sensor data is vulnerable to RAs given that the attacker can easily learn a sequence of previously generated data and repeat it. Similarly, a meter unprotected with firewall that is connected to the Internet will be vulnerable to data injection attacks.

Risk management uses vulnerability assessment results and combines them with an estimation of the effect that a vulnerability can have on the SG. Risk management in SG includes two key tasks: 
\begin{enumerate}
\item{\emph{Contingency analysis:} assessing the effect that the loss of a component such as a transmission line, generator, transformer, or sensor, can have on the dynamic stability and operating state of the grid.} 
\item{\emph{Cascading failures analysis:} analyzing the propagation of failures over the grid.}
\end{enumerate} 

The later requires understanding the interdependencies between the various grid elements to anticipate the cascading chain of events that may occur when some components are lost. For example, the loss of a transmission line in heavy loaded conditions might lead to cascading failures resulting in a blackout while the loss of another line might have unnoticeable effects.

Vulnerability assessment and risk management are continuously evolving processes using which the operator continuously learns about potential threats and vulnerabilities so as to improve its protection of the system.
}
\item{\emph{Security reinforcement:} once threats and their associated effects are characterized, policies for reinforcing the grid security must be derived. However, such reinforcement procedures are typically subject to budgetary and investment constraints. Thus, the results obtained from vulnerability and risk management can be used to create a ranking of the most critical components to protect first. Security reinforcement can include encrypting a number of sensor readings, replacing some meters with more sophisticated and capable models~\cite{SanjabSaadDataInjTSG}, 
%ones which can run computationally demanding cryptographic algorithms, 
replacing wireless with wired communication, or implementing robust control designs. %which can withstand disturbances, among others. 
Security reinforcement is a robustness measure which aims at securing the grid against a range of potential threats.}
\end{enumerate}
%Such vulnerability assessment and reinforcement techniques have been implemented in~\cite{SanjabSaadDataInjTSG} to secure the grid against data injection attacks targeting energy markets. Moreover, the work in~\cite{SanjabSaadCPSWeek} proposes reinforcement strategies to face data injection attacks targeting wide area protection schemes.       
\subsection{Scanning and Detection}\label{subsec:Detection}
Vulnerability assessment, risk management, and security reinforcement constitute a preventive measure to thwart potential attacks which can target the grid. However, they cannot completely secure the grid against all types of attacks for two main reasons: 
\begin{enumerate}
\item{\emph{Undiscovered threats:} there are many emerging vulnerabilities and attack strategies of which the operator might not be aware.}% and which a very skilled adversary can exploit.}
\item{\emph{Budget constraints:} budgetary constraints can limit the volume of possible security reinforcement implementations.} 
\end{enumerate}

Therefore, the operator must continuously scan the system to detect new threats which have passed the attack prevention defense lines. 
This is crucial for stealthy and advanced persistent threats which penetrate the system and run long-term stealthy attacks that cannot be obviously identified. %do not have obvious effects on the physical system such as manipulating electricity prices~\cite{SanjabSaadDataInjTSG}.
%The goal of scanning and detection falls in one of two categories based on the attack types: 
%\begin{enumerate}
%\item{\emph{Apparent attacks:} for attacks causing obvious physical damage to the system, such as ones affecting the stability and controllability of the grid~\cite{NERCUkranian,Stuxnet,SniperSGAttack}, the goal is to identify the attack mechanism, its source, and functionality to be able to subsequently stop and eliminate it.}
%\item{\emph{Stealthy attacks:} for stealthy persistent attacks which can target the system for a long period of time without showing attack symptoms such as replay attacks~\cite{AttDetDorfler,ReplayAttacksSG}, stealthy data injection attacks~\cite{AttDetDorfler,SanjabSaadDataInjTSG}, the goal is to continuously scan the system to identify the persistent unnoticeable attacks and subsequently eliminate their effect on the grid's operation.} 
%\end{enumerate}

Various detection mechanisms have been studied in literature with each focusing on detecting a type of SG attacks. For example, the work in~\cite{MaliciousDataDetectionIllinois} provides a methodology for detecting stealthy data injection attacks targeting the state estimator %, through observing the perturbation that data injection introduces to the estimation of system parameters. %knowledge of system parameters to characterize the attacked measurements. 
%In addition, 
while the authors in~\cite{DetectingTimeSynchroAtt} propose a detection mechanism against TSAs targeting PMUs. Moreover, the authors in~\cite{ReplayAttacksSG} propose a detection mechanism against replay attacks.  %%%%%%%%%%%%%%%%the last can be potentially removed for space
 
\subsection{Mitigation, Elimination, and Restoration}\label{subsec:MitigationResilienceRestoration}
Once an attack is detected using detection mechanisms or by witnessing an apparent damage to the grid, mitigating the effect of the attack and eliminating it become essential for restoring the normal operating state. 
\begin{enumerate}
\item{\emph{Mitigation:} during the occurrence of an attack, mitigation measures can be deployed to reduce its effect on the system. Mitigation techniques may include: a) power system protection techniques which can be used for stopping the propagation of disturbances, b) spinning reserves or distributed generation which can be leveraged to meet the loss of generation, c) load shedding which disconnects a part of the total load in order to prevent a complete collapse of the system, and d) islanding which splits the grid into small disjoint systems to stop the propagation of failures and dynamic disturbances. %, and e) the use of high-voltage direct current (HVDC) transmission lines to prevent inter-area disturbance propagation.}
Mitigation techniques are forms of resilience measures in which the system sacrifices some operational quality to decrease the risk of a complete collapse.} 
%A resilient design of the grid can mitigate extreme attacks by sacrificing  (shedding load, disconnectingwhen faced with an attack to meet minimal critical performance requirements until the event is eliminated. %The work in~\cite{} provides an analysis of resilience of smart grids in which resilience is compared to a reed which bends when faced by a storm which helps it survive as compared to an oak tree which, even though is robust, lacks flexibility and brakes under heavy winds. 
%Such a flexibility can lead to the survivability of the grid under attack. For example, aiming at meeting $100\%$ of the system load when faced with an attack might put too much stress on generators, transmission lines and transformers leading to their failures which can lead to a blackout. However, deciding to meet only the critical load (a certain proportion of the total load) reduces the stress on the system and gives it flexibility to survive the attack.  

%A resilience measure to mitigate the effect of attacks can make use of distributed generation, such a home-scale renewable energy resources and storage units, to meet critical loads in their corresponding neighboring areas in the case of interruption of service, or load shedding, due to an attack.}
\item{\emph{Elimination and Restoration:} to restore the normal operating state of the system once an attack is detected, taking prompt actions to eliminate it becomes critical. Elimination can be done by various means such as replacing compromised components or updating their software. After threat elimination, SG elements that had been disconnected such as transmission lines, generators, and loads can be reconnected to restore normal operation.}

%
%\item{\emph{Restoration:} restoration corresponds to bringing back the system to its normal state of operation after the attack has been eliminated. Restoration requires replacement of the damaged equipment, as well as reconnecting generators which have been disconnected and load which have been shed during the attack period.}
\end{enumerate}

\begin{figure*}[t]
\centering
\begin{table}[H]
\centering
\caption{{\large Summary of SG Security threats, Challenges, and Solutions.}}
\label{tab:Summary}
\begin{tabular}{|c|l|l|}
\hline
\multirow{9}{*}{\textbf{Threats}}    & \multirow{2}{*}{Integrity attacks} & Data injection attacks. \\ \cline{3-3} 
                            &                          &Time synchronisation attacks. \\ \cline{2-3} 
                            & Availability attacks & Denial of service attacks. \\ \cline{2-3} 
                            & \multirow{3}{*}{Dynamic system attacks} & Replay attacks. \\ \cline{3-3} 
                            &                          & Dynamic data injection attacks. \\ \cline{3-3} 
                            &                          & Covert attacks.\\ \cline{2-3} 
                            & \multirow{2}{*}{Physical attacks} & Physical damage. \\ \cline{3-3} 
                            &                          & Meter manipulation. \\ \cline{2-3} 
                            & Coordinated attacks & Multiple concurrent types and targets. \\ \hline
\multirow{6}{*}{\textbf{Challenges}} & \multirow{2}{*}{Limitation of cyber security solutions}      & Presence of physical system.  \\ \cline{3-3} 
                            &                          & Different risk propagation mechanisms.  \\ \cline{2-3} 
                            &Limitation of reliability evaluation solutions                       & Presence of unlikely coordinated failures. \\ \cline{2-3} 
                            & Limitation of conventional control-theoretic solutions                      & Presence of cyber layer.  \\ \cline{2-3} 
                            & Tradedoff: security vs. performance                       & Security must preserve operational requirements. \\ \cline{2-3} 
                            & Aging components                       & Old equipments: need security patching.  \\ \hline
\multirow{7}{*}{\textbf{Solutions}}  & \multirow{2}{*}{Prevention}      & Vulnerability assessment and risk analysis.   \\ \cline{3-3} 
                            &                          & Security reinforcement.  \\ \cline{2-3} 
                            & \multirow{2}{*}{Detection}      & Focus: stealthy and persistent threats.  \\ \cline{3-3} 
                            &                          & Constraints: budget.  \\ \cline{2-3} 
                            & \multirow{3}{*}{Mitigation, elimination, and restoration}      & Mitigation: block propagation, leverage reserves. \\ \cline{3-3} 
                            &                          & Elimination: Update compromised components. \\ \cline{3-3} 
                            &                          & Restoration: reconnect disconnected components. \\ \hline
\end{tabular}
\end{table}
%\\\bottomrule %\nonumber
\end{figure*}

After an attack, the operator's knowledge about this attack and the vulnerabilities of the system as well as the ways of detecting and mitigating this attack evolves. Therefore, this allows for the operator to update its defense policies to improve the security of the grid.

Table~\ref{tab:Summary} summarizes the discussed threats, challenges, and solutions while providing relevant comments.
To successfully analyze SG security and deploy the aforementioned solutions, a number of analytical frameworks can be leveraged, as discussed next.
%As has been discussed in Section~\ref{sec:Challenges}, various challenges face the implementation of security measures. To this end, analytical tools which can help in the implementation of the discussed defense mechanism are introduced next. 
   
\section{Analytical Frameworks for Smart Grid Security Analysis}\label{sec:AnalyticalTools}
Implementing effective SG security solutions requires analytical frameworks which enable modeling of the grid's cyber and physical systems and their tight coupling, the interdependency between various grid components, and the decision making processes of the operator and attackers. %propagation of attacks and failures throughout the grid.
%the networked cyber-physical interconnectivity of the SG and model the decision making processes    

%As discussed in Section~\ref{sec:Challenges}, even though information security, reliability evaluation, and control-theoretic techniques show some limitations when it comes to solving SG security problems, combining the improvements that each of these fields have introduced throughout the years is extremely useful to improving the security of the grid. 
%To this end, SG security has emerged as an interdisciplinary field in which experts from power systems, information security, communication systems, mathematics, reliability evaluation, control systems, psychology, and criminology, cooperate to propose holistic security solutions. 
%
%One key aspect in the validity of any proposed security solution lies in modeling both the cyber and physical systems of the grid, their tight coupling, the interdependency between their various components, and the propagation of attacks and failures throughout the grid. In fact, threats targeting the SG are of cyber-physical nature and hence require \emph{cyber-physical security solutions}.
%
In addition to using solutions from information security, power system protection, control theory, and reliability evaluation, additional analytical tools are very useful in modeling and studying SG security problems as discussed next.  
\subsection{Modeling Using Networked Control Systems}\label{subsec:NetworkedControl}
Networked control systems (NCSs)~\cite{NetCtrlSysSurvey} combine communication and information technologies with control system designs to model CPSs such as the SG. Indeed, in NCSs, a shared communication network is responsible for the communication between the various sensors, actuators, and controllers in the CPS. 
As a result, NCSs can be used to model the cyber-physical nature of the smart grid which is extremely important for studying potential threats and deriving appropriate security solutions for the SG. %cyber-physical security solutions. 

%the power grid model interconnected dynamic systems while accounting for communication technologies (delays, channel modeling, packet rates/dropouts   
\subsection{Graph-Theoretic Techniques}\label{subsec:GraphTheory}
Given that the SG is a networked cyber-physical system (NCPS), \emph{graph-theoretic} techniques can be useful to model the network interconnectivity between the grid components. 
In fact, a graph is represented by a set of vertices and a set of edges connecting these verticies. For SG security applications, the verticies can represent components such as generators, transformers, loads, or meters. %, relays, antennas, transmitters, and receivers. 
while edges can model the interconnectivity between these components. The modeling of this interconnectivity can consider real, physical connectivity between the components~\cite{GTDataInjProtection} or functional, logical connectivity modeling the interdependencies between those components~\cite{SanjabSaadCPSWeek}. 

Hence, graph-theoretic methods are very useful for understanding the networked nature of the SG and analyzing the interconnectivity between its elements to shed light on the threats facing the grid and their propagation. %For example, the work in~\cite{GTDataInjProtection} uses graph-theoretic analysis to propose a defense mechanism against data injection attacks on the state estimator.   
\subsection{Game-Theoretic Techniques}\label{subsec:GameTheory} 
\emph{Game theory} is a set of mathematical tools used to analyze strategic interaction and decision making between entities, referred to as players, with interconnected, conflicting or aligned, interests. In a typical SG security setting, an attacker aims to choose an attack strategy to maximize the damage caused to the grid while the operator (defender) aims at choosing a defense strategy to minimize the damage to the system. Thus, due to the conflicting objectives of the attackers and defenders, game-theoretic techniques~\cite{NetworkSecGTAlpcanBasar} provide invaluable tools to model their optimal decision making and hence finding the best defense strategy given the potential strategies of the attackers. 

Various works have applied game-theoretic techniques to the analysis of smart grid security problems. For example, our recent work in~\cite{SanjabSaadDataInjTSG} used game theory to characterize the best set of meters to defend in order to thwart data injection attacks which can be carried out by multiple adversaries.

\section{Future Outlook}\label{sec:Conclusion}
%The cyber-physical coupling in the smart grid plays a very important role in improving the efficiency and sustainability of the grid. However, it poses various security challenges. As a result, understanding the threats that can target the grid is of utmost importance to derive valid security solutions. To this end, this paper has explored the cyber-physical threats that can target the smart grid while investigating their effects on the stability and operation of the system. Moreover, the challenges of studying smart grid security problems and devising appropriate solutions have been highlighted and analyzed. In addition, solution approaches proposed in literature have been systematically analyzed.  
%
%Various research efforts have proposed valuable approaches for solving the multiple security threats targeting the smart grid. The key advancements needed in future research works is to model the smart %grid as a networked control system which enables detailed modeling of both the cyber and physical realms of the grid and their interconnectivity. 
%
%. Moreover, mathematical tools from graph theory and game theory can be invaluable in modeling the networked nature of the grid and the decision making processes of the defenders and adversaries. Such models will 
%
%Such models would allow for the derivation of appropriate security solutions for the cyber-physical smart grid to thwart, detect and mitigate the culminating cyber-physical threats.    

In this paper, a comprehensive analysis of the various SG security threats, incurred challenges, and potential solutions have been investigated. This analysis paves the way for more in-depth investigations of each of the discussed threats and their correlations in order to quantize their combined effects on the SG. This will enable devising appropriate solutions against coordinated attacks which threaten the sustainability of current and future smart grids.          
\bibliographystyle{IEEEtran}
\bibliography{reference}  
\end{document}